\documentclass{myart}
\usepackage{epsfig,times}
\begin{document}
\begin{frontmatter}
\title{Calibration of the STAR Forward Time Projection Chamber with
Krypton-83m}
\collab{V.~Eckardt, T.~Eggert, H.~Fessler, H.~H\"ummler, G.~Lo~Curto,
M.~Oldenburg, N.~Schmitz, A.~Sch\"uttauf, J.~Seyboth, P.~Seyboth, M.~Vidal}

\address{Max-Planck-Institut f\"ur Physik, F\"ohringer Ring 6, 80805
M\"unchen, Germany}

\begin{abstract}
The principles of the calibration of a time projection chamber with
radioactive Krypton-83 are explained. The calculation of gain
correction factors and the methods of obtaining a precise energy
calibration are illustrated. The properties and advantages of
\nuc{83m}{Kr} are summarized and compared to other radioactive
calibration sources. It was shown that the Krypton calibration for the
STAR FTPC is feasible and recommendable although the pad geometry of
the FTPC causes a considerable deterioration of the measured spectrum
due to only partially detected charge clusters.
\end{abstract}

\end{frontmatter}

The STAR experiment (Solenoidal Tracker at RHIC) is one of four
experiments operating at the Relativistic Heavy Ion Collider (RHIC) in
Brookhaven. STAR searches for hadronic signatures of the quark gluon
plasma formation and investigates the behavior of strongly interacting
matter at high energy density in collisions of heavy nuclei
\cite{star,Harris,Wieman}. While the tracking in the central region is
provided by a large TPC, the two Forward Time Projection Chambers
(FTPCs) are required to give position, charge, and momentum
information of particle tracks in the forward rapidity regions of
the STAR experiment \cite{Proposal,Schuettauf}.

The calibration method using Krypton-83m has been developed by the
ALEPH collaboration \cite{aleph} and is being used by the DELPHI
experiment \cite{delphi}. It is also applied very successfully by the
NA\,49 experiment where the achieved measured accuracy (i.\,e. the relative
error of the mean value $\Delta E/E$) is better than 0.5\,\%
\cite{na49nim}.  

\section{The Calibration}

A calibration of a Time Projection Chamber (TPC) is performed for two
main purposes: The {\em gain correction}, which is needed to
compensate variations in the front-end electronics (FEE) and in the
gas gain, and the {\em energy calibration}, which is performed in
order to determine the amount of electric charge that is measured at a
known energy deposition in the chamber.

\subsection{Gain Correction}

The gain of the pre-amplifier electronics varies from chip to chip
(see also section~\ref{measurements}) and within a chip from channel
to channel. Moreover, variations in gas gain may occur due to
inhomogeneties of the electric field and geometrical non-uniformities.

A straightforward and effortless relative gain correction for the
pre-amplifier electronics can be perfomed by a so-called pulser
calibration. This is done by injecting calibrated charge pulses into
the anode or gating grid wires of the multi-wire proportional chambers
(readout chambers).\label{pulsercalib} Each pulse induces a charge on
each cathode pad $i$, resulting in a charge signal $q_i$ which is
proportional to the channel's response. Multiplicative calibration
constants $c_i$ are then obtained by dividing the average charge per
pad
\begin{equation}\bar{q}=\frac{1}{N} \sum_{i=1}^N q_i\end{equation}
($N=$ number of pads) by the charge induced on an individual pad:
\begin{equation}c_i=\frac{\bar{q}}{q_i}\end{equation}

Each measured charge per pad $q^{\mathrm{meas}}_i$ (i.\,e. integrated
ADC channels) then gets multiplied by this calibration constant in
order to obtain the calibrated charge:
\begin{equation}q^{\mathrm{cal}}_i=c_i q^{\mathrm{meas}}_i\end{equation}

However, pulsing the anode wires induces a signal on all pads
simultaneously, producing a large current load on each FEE channel
which may result in a modification of the electronics response. Thus,
the pulser calibration can only be considered as a first step in the
calibration procedure.

\subsection{Energy Calibration}

It is not possible to perform an absolute energy calibration by a
pulser calibration only. Therefore, an alternative calibration method
using a radioactive source has to be considered.  Although particle
identification via $\d E/\d x$ measurement is difficult with the STAR
FTPC \cite[Sect.\,6.6]{Proposal}, an energy calibration is useful to
optimize the accuracy of the position measurement. It can easily be
done with a radioactive source such as \nuc{83m}{Kr} which is injected
into the chamber gas.

An electron emitted during the decay process of \nuc{83m}{Kr} (see
section~\ref{decayprocess}) produces secondary electrons by ionization
in the gas, where the number of produced electrons is proportional to
the initial decay energy (i.\,e. the kinetic energy of the primary
electron). Since the decay energies of suitable sources are relatively
low ($E \ll m_e c^2$), the primary electron is stopped very quickly,
and the secondary electrons emerge from virtually one point. These
secondary electrons are then drifted by an electric drift field to the
readout chambers where they are amplified at the anode wires, and the
charge signal from the gas amplification is detected at the cathode
pads.

The decay energies of the radioactive source are known, and thus the
corresponding energies can be assigned to the peaks of the measured
charge spectra. Starting from a distinct reference peak, the energy
that corresponds to a measured charge $q$ is then calculated by
\begin{equation}
E(q)=q \; \frac{E_{\mathrm{ref}}}{q_{\mathrm{ref}}}
\end{equation}
where $E_{\mathrm{ref}}$ is the energy and $q_{\mathrm{ref}}$ is the
measured charge of the reference peak. However, because the highest
decay energy of \nuc{83m}{Kr} is relatively high compared to the
energies of a few keV deposited by a minimum ionizing particle, the
anode voltage $U$ has to be reduced from the value suitable for
minimum ionization. The results from the energy calibration must then
be extrapolated to higher gain voltages. As the dependence of the
amplification on the anode voltage has been measured, such an extrapolation
is accurate.

Further applications of the Krypton calibration are a cross-check of
the pulser calibration and an investigation of gas flow effects in the
chamber by studying the location of Krypton decays at the beginning of
the injection.  Moreover, a calibration is needed to check the
detector linearity over a wide energy range, and to monitor the long
term detector stability.

\section{The STAR Forward TPC}

The high track density in the forward rapidity regions of the STAR
experiment covered by the FTPCs (pseudorapidity range of
$2.5<|\eta|<4$) requires a special design for the tracking detectors
in these regions.  In contrast to conventional time projection
chambers, the two Forward TPCs of the STAR detector use radial drift
fields, and the curved readout chambers are part of the outer cylinder
walls. The compactness of the FTPCs (60\,cm diameter and 120\,cm length)
with only 22 cm drift length 
and the use of a cool gas provides a very good position resolution (150
$\mu$m) and two-track separation (1 mm) \cite{Marst,Hummler}. In order to
minimize the number of readout channels without compromising the detector
performance the surface is covered only partially by cathode pads. There
are 10 padrows per FTPC, with 960 pads each. The length of one pad is
20\,mm (in $z$ direction, where $z$ points along the beam pipe), and the
pad pitch is 1.9\,mm (in azimuthal direction $\phi$). The distance between
the padrows is intermittently 65\,mm and 85 mm. For further details on the
STAR FTPC, see \cite{Proposal,Schuettauf}.

\begin{figure}
\mbox{\epsfig{file=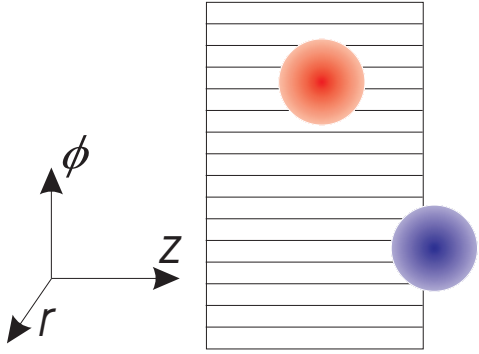,width=0.45\linewidth}}
\hfill
\parbox[b]{.45\textwidth}{\caption{\label{padrow} Two electron clusters
produced by Krypton decays and a row of cathode pads. As the padrows
are not adjacent to each other, for some of the clusters the charge
can only be partially detected \cite{Marst}.}}
\end{figure}

As the padrows are not adjacent to each other, a problem arises when
localized charge clusters from radioactive decays are to be
detected. As illustrated in figure~\ref{padrow}, only a fraction of
the cluster charge can be detected by the pads when the cluster is
located at the border of a padrow. The question whether a calibration
with a radioactive source can nevertheless be done triggered a
systematic investigation on the practicability of a calibration with
radioactive Krypton.

\begin{figure}[b]
\mbox{\epsfig{file=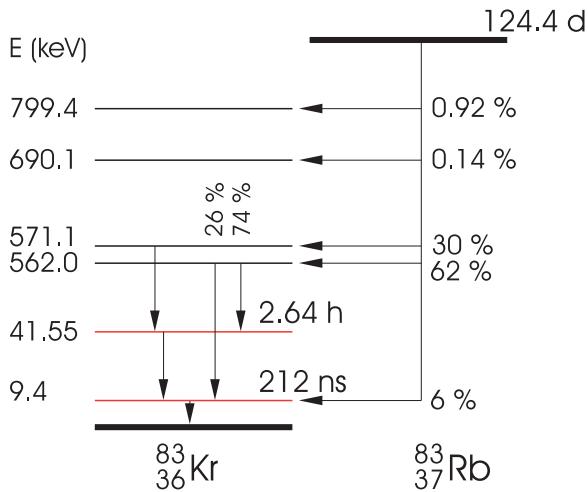,width=0.55\linewidth}}
\hfill
\parbox[b]{.35\textwidth}{\caption{\label{rbdecay} Decay scheme of 
\nuc{83}{Rb} $\rightarrow$ \nuc{83}{Kr}. The ground state of $^{83}$Rb
decays predominantly (76\,\%) to the isomeric excited state
\nuc{83m}{Kr} at $E=41.6$\,keV. This most important level of
\nuc{83}{Kr} for calibration is populated through the 571\,keV and
562\,keV intermediate levels. It decays entirely to the 9.4\,keV
state, which then decays to the \nuc{83}{Kr} ground state.}}
\end{figure}

\section{Properties of Krypton-83} \label{decayprocess}

\nuc{83}{Kr} is a stable isotope produced from \nuc{83}{Rb} which decays by
electron capture with a mean lifetime of 124 days. The decay scheme
\cite{Lasiuk} is shown in figure~\ref{rbdecay}. The ground state of
\nuc{83}{Kr} is not reached directly in the \nuc{83}{Rb} decay but via
nuclear transitions from intermediate excited \nuc{83}{Kr} levels. The
relevant level for drift chamber calibration is the isomeric
(metastable) 41.6\,keV state \nuc{83m}{Kr}. It is fed down from two
higher \nuc{83}{Kr} levels in 76\,\% of all \nuc{83}{Rb} decays and
has a lifetime long enough ($\tau = 2.64$\,h) for the \nuc{83m}{Kr} to
be introduced and distributed in the drift chamber before it
decays. The isomeric state \nuc{83m}{Kr} decays entirely by an E3
nuclear de-excitation transition to the 9.4\,keV level with a
transition energy $E_{\mathrm{tr}}$ of 32.2\,keV. This short-lived
state then decays immediately ($\tau =212$\,ns) to the \nuc{83}{Kr}
ground state with $E_{\mathrm{tr}}$ = 9.4\,keV. In contrast to other
common calibration sources like \nuc{55}{Fe} or \nuc{57}{Co}, almost
no low energy photons are produced since both transitions from the
41.6\,keV and the 9.4\,keV states proceed almost entirely by internal
conversion (IC) with ratios of electron-to-photon emission of
$e^-/\gamma \approx 2 \cdot 10^3$ and $e^-/\gamma \approx 20$
respectively.

The kinetic energy $E_{\mathrm{e}}$ of an IC electron is given by
\vspace{-1mm}
\begin{equation} \label{kinetic}
E_{\mathrm{e}} = E_{\mathrm{tr}} - E_{\mathrm{b}}
\end{equation}
\vspace{-4mm}
where $E_{\mathrm{b}}$ is the atomic binding energy of the
electron. The subsequent atomic de-excitation of the electron hole
produces either X-rays or Auger electrons with an energy equal to $E_{\mathrm{b}}$.

For the 41.6\,keV-to-9.4\,keV transition the internal $\gamma$
conversion occurs predominantly on an electron in an outer (L, M, N) shell,
the total electron emission probability from such shells being 77\,\%
\cite{Lasiuk}. The binding energies for the outer shells are small $(E_{\mathrm{b}}
\le \, 1.9 \, {\rm\,keV})$. Furthermore, the de-excitation of an
outer-shell hole leads in most cases to the emission of an Auger electron
rather than an X-ray. Therefore practically the full energy $E_{\mathrm{tr}}$
of 32.2\,keV is carried away by electrons and can be collected. If, however,
the internal conversion occurs in the K shell (23\,\% of the cases) an IC
electron with an energy $E_{\mathrm{e}}$ of 17.8, 18.1, 19.5 or 19.6\,keV is
produced, corresponding to four K sub-levels with binding energies of
14.3, 14.1, 12.7 and 12.6\,keV \cite{Lasiuk}, respectively, according to
eq. (\ref{kinetic}). In addition, the de-excitation of the K-shell hole
yields X-rays or Auger electrons with an energy equal to the binding
energy. 

For the 9.4\,keV-to-ground state transition the low transition
energy allows only internal conversion on an outer-shell electron, so that
practically the full transition energy of 9.4\,keV is carried away by the IC
electron (95\,\% of the cases) or by the escaping $\gamma$ (5\,\%).

For more details on the released energies and on the branching fractions,
see \cite{Lasiuk}.

The overall decay spectrum has been simulated, using the branching
fractions in table 4 of \cite{Lasiuk} and assuming a resolution of
$\sigma/E = 6\,\%$, where $\sigma$ is the width of a Gaussian. The simulated
spectrum is shown in figure~\ref{krsim1}; it is composed  of five parts:
\begin{figure}
\mbox{\epsfig{file=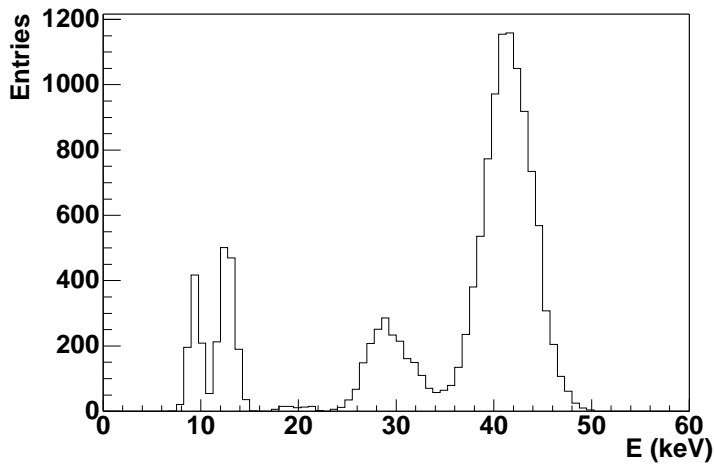,width=0.7\linewidth}}
\hfill
\parbox[b]{.29\textwidth}{
\caption{\label{krsim1}
Simulated decay spectrum of \nuc{83m}{Kr} with an assumed line width
of $\sigma/E=$\,6\,\%. Noise is neglected.}}
\end{figure}
\vspace{-3mm}
\begin{itemize}
\item[$\bullet$]
The strong peak at 41.6\,keV is due to the electrons from the dominant
modes of the two-step decay of the 41.6\,keV isomeric state via the 9.4\,keV
state. Because of the short lifetime of the 9.4\,keV state the two decay
energies of 32.2\,keV and 9.4\,keV are deposited in the majority of cases at
the same location such that the total decay energy is collected in a single
cluster. 
\item[$\bullet$]
The smaller peak around 30\,keV comes from two contributions:
(a) electrons with $E_{\mathrm{e}}$ = 32.2\,keV from the 41.6\,keV-to-9.4\,keV
transition when the 9.4\,keV $\gamma$ escapes; (b) K-shell IC
electrons ($E_{\mathrm{e}} \, \approx$ 19.5\,keV) together with
IC electrons from the decay of the 9.4\,keV state at the
same location, leading to a total energy deposition of $\sim$ 29\,keV.
\item[$\bullet$]
The very small accumulation around 20\,keV is due to K-shell IC electrons
($E_{\mathrm{e}} \approx$ 19.5\,keV) when the 9.4\,keV $\gamma$ escapes.
\item[$\bullet$]
The peak at 12.7\,keV is produced by the conversion of K-shell X-rays in
the chamber gas, away from the location of their origin.
\item[$\bullet$]
The peak at 9.4\,keV results from either the conversion of the escaped 9.4
keV $\gamma$ in the gas or the 9.4\,keV electron energy, separated from the
preceding 32.2\,keV transition.
\end{itemize}

\begin{figure}
\mbox{\epsfig{file=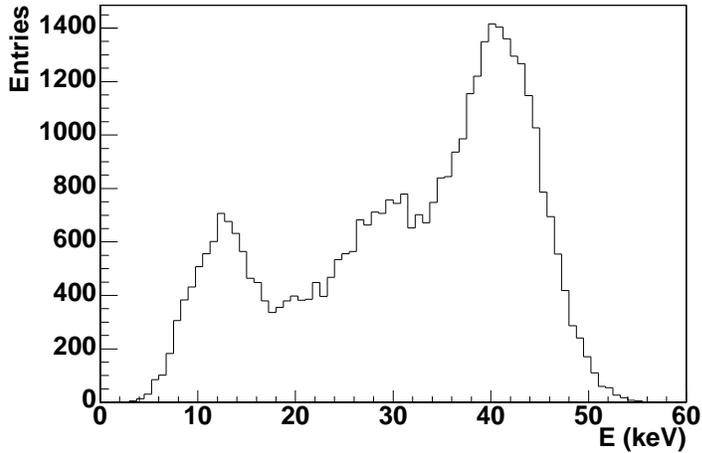,width=0.7\linewidth}}
\hfill
\parbox[b]{.29\textwidth}{
\caption{\label{krsim2} Simulation of the \nuc{83m}{Kr} decay
spectrum as seen by the two padrows of the test-setup.  Diffusion,
partially detected clusters, and a threshold cut are taken into
account.}
}
\end{figure}

Figure~\ref{krsim2} shows a more detailed simulation, where partially
detected clusters caused by the pad geometry, diffusion, the detector
resolution, and cut criteria comparable to those in the experiment
were considered. The 41.6\,keV peak is clearly visible while the
several peaks around 30\,keV as well as the 9.4\,keV and the 12.7\,keV
peaks are hardly separable. Moreover, the shape of the low energy part
of the spectrum is strongly affected by the chosen threshold.

The isomeric state \nuc{83m}{Kr} is a very useful isotope for
calibrating a time projection chamber for several reasons: in contrast
to other frequently used calibration sources such as \nuc{55}{Fe},
\nuc{83}{Kr} is a gas which can be distributed over the chamber volume
using the existing gas system. Therefore, no laborious unmounting of
the chamber and installation of Fe sources is necessary. The mean
lifetime of \nuc{83m}{Kr} is short enough to ensure the chamber to
operate normally again after a reasonably short time, i.\,e. once a
few half-lives have passed after cutting off the Krypton supply to the
chamber. If required, the gas can be left within the closed gas system
until the radioactivity subsides. On the other hand, the mean lifetime
of the isomeric state is long enough for a sufficient number of subsequent
decays to occur inside the chamber. Krypton's parent isotope $^{83}$Rb has
a sufficiently long mean lifetime ($\tau = 124$\,d), is a solid, and thus
may be mounted as a foil inside a bypass line of the gas system.

As it is not possible to trigger on the Krypton decays, a random
trigger must be used. The decay rate must be high enough to ensure
good statistics within a reasonable time (an estimate of the number of
events needed is carried out at the end of section~\ref{measurements}).
Due to random triggering there is no information on the decay position
in the drift direction. Therefore, to obtain a clean spectrum,
absorption (mainly caused by oxygen pollution in the gas) must be
small\footnote{At an oxygen concentration of 5\,ppm and the maximum
drift time of 50\,$\mu$s, 17\,\% of the electrons are
absorbed \cite{Eggert}.}.

\section{The Measurements} \label{measurements}

As neither the STAR FTPC nor the FTPC readout electronics were
completed when the measurements were performed, a small test-setup
using the NA\,49 readout electronics \cite{Rauch,Bieser} was
constructed. For this test-setup, the same materials and gas-mixture
(Ar/CO$_2$ 50/50) as for the FTPC, and a pad geometry comparable to
the FTPC (two non-adjacent padrows with 16 pads each) were used. The
major differences between the test-setup and the FTPC was the axial
drift field (in contrast to the radial drift field of the FTPC) and
the planar readout chamber. However, the charge deposited on the pads
is hardly affected by the drift field geometry and the cluster sizes
were comparable to those measured in a test TPC with a radial drift
field. For details on the test-setup and the measurements, see
\cite{Eggert}.

The calibration was performed in three steps: the pulser calibration,
the calibration with \nuc{55}{Fe}, and the calibration with \nuc{83m}{Kr}.

\subsection{The Preparatory Calibrations}

The pulser calibration is described earlier in
section~\ref{pulsercalib}. The amplification factors of the 16
channels per preamplifier/shaper chip decrease characteristically with
increasing channel number $i$; they differ by up
to 15\,\%. This behavior is a property of the architecture of the
NA\,49 FEE chips and is not observed in the FTPC electronics where the
fluctuation of the amplification factor is below 3\,\%. Thus, a
pad-wise calibration was necessary for the test-setup but might not be
needed for the calibration of the FTPC. However, the pulse shape of
the FTPC electronics is affected by the current load on the chip. The rise
time of the pulse is reduced as the number of pulsed channels increases
\cite{sn324} which will affect the total integrated cluster charge. Thus, a
cross-check of the pulser calibration has to be performed.

\begin{figure}
\mbox{\epsfig{file=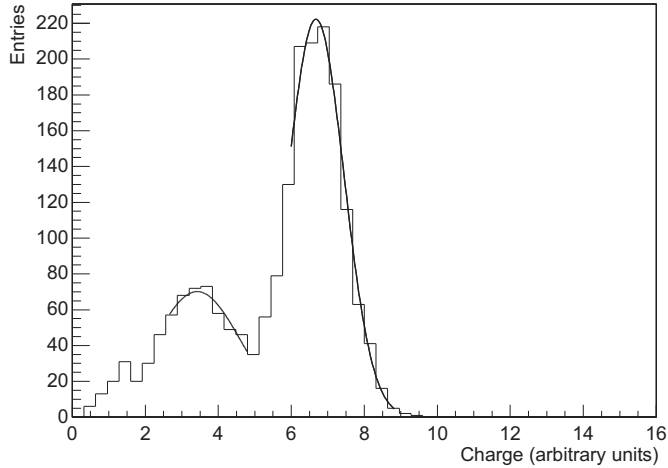,width=0.7\linewidth}}
\hfill
\parbox[b]{.29\textwidth}{\caption{\label{fespec} A Fe spectrum
measured on a single pad. A combined fit
consisting of two Gaussians was performed and the two separate
Gaussians are shown in this plot.}}
\end{figure}

For the measurements with the \nuc{55}{Fe} source, two Aluminum strips
doped with \nuc{55}{Fe} were mounted at the cathode plate of the drift
field cage, opposite to the two pad\-rows. \nuc{55}{Fe} decays via
electron capture to $^{55}$Mn which subsequently emits an X-ray photon
from the K-shell with an energy of 6\,keV. This photon may ionize an
Argon atom in the gas which releases an electron from the K-shell,
which has a binding energy of 3.2\,keV. In addition, either Auger
electrons or an escaping X-ray photon with an energy of 3.2\,keV are
emitted, the latter being responsible for the 2.8\,keV escape
peak \cite{Eggert}. The supplementary calibration using \nuc{55}{Fe}
was performed for several reasons:

\begin{itemize}
\item A cross-check of the pulser calibration and a first order energy
calibration was needed to tune the anode voltage for the Krypton
measurements.
\item Determining the widths of the 6\,keV main peak and of the 3\,keV
escape peak gave an accurate measurement of the chamber resolution.
\item A comparison of the gas amplification at anode voltages in the
operating range of the FTPC (1\,700\,V to 1\,800\,V) was required.
\item The clear Fe spectrum allowed a straightforward determination of
suitable cut criteria in order to reduce noise.
\end{itemize}

The electron energy deposition of \nuc{55}{Fe} roughly corresponds to
the energy deposition of a minimally ionizing particle. Thus, it was
possible to perform an energy calibration at the FTPC's operating
anode voltage of $U=$~1\,750\,V. In addition, further measurements at
voltages from 1\,700\,V to 1\,800\,V were carried out. These results
were extrapolated to lower voltages in order to find the anode voltage
where the complete Krypton spectrum can be measured without electronics
saturation caused by the high decay energies of \nuc{83m}{Kr}.

\begin{figure}
\mbox{\epsfig{file=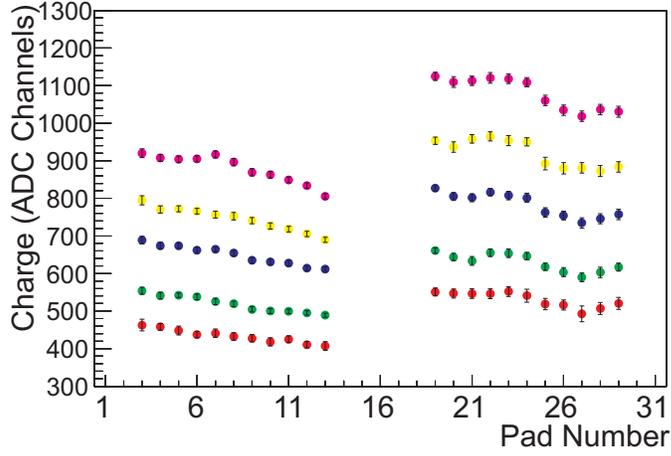,width=0.63\linewidth}}
\hfill
\parbox[b]{.29\textwidth}{
\caption{\label{feladung}
The measured charge $q_i$ (integrated ADC channels) from the 6\,keV Fe
decays of each pad $i$ without any calibration. Anode voltages from the
bottom up: $U=$ 1\,700\,V, 1\,725\,V, 1\,750\,V, 1\,775\,V,
1\,800\,V. }}
\end{figure} 

\begin{figure}
\mbox{\epsfig{file=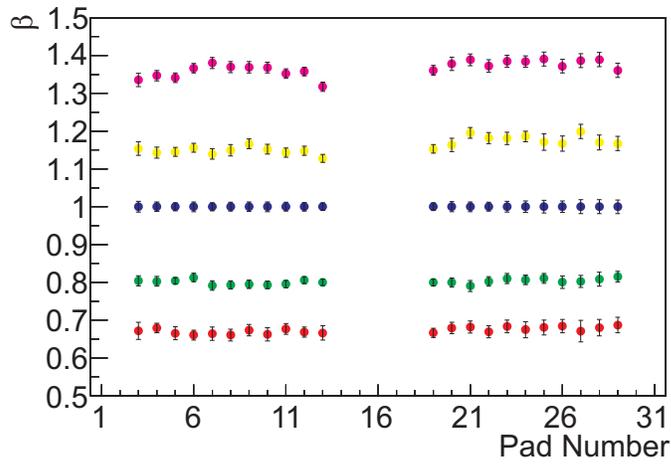,width=0.63\linewidth}} \hfill
\parbox[b]{.29\textwidth}{
\caption{\label{fekaladung} The measured charge $q_i$ from Fe decays of
each pad $i$ (as in fig.~\ref{feladung}) with calibration factors obtained
by the reference measurement at
1\,750\,V. $\beta=q_i(U)/q_i$(1\,750\,V)}}
\end{figure}

For each pad a separate Fe spectrum was measured and a combined fit of
the two peaks was performed (see figure~\ref{fespec}). For the
calculation of the calibration factors only the position of the 6\,keV
peak was considered. The width of the peak and hence the detector
resolution were determined to be $\sigma/E \approx$\,12\,\%.  The
positions of the fitted Gaussians on the charge spectra of each pad
(except the marginal pads) is shown as a function of the pad number in
figure~\ref{feladung}.  By applying the Fe calibration factors from a
reference measurement at 1\,750\,V to measurements at different anode
voltages, the consistency of the Fe calibration and thus the expected
precision for the Krypton calibration was investigated (see
figure~\ref{fekaladung}, where only statistical errors are shown). The
resulting uncertainty is determined to be $<$\,2\,\%.

The comparison of the calibration factors obtained with the pulser and
the iron source shows notable differences. The drop in amplification
with increasing channel number is smaller for calibration using the Fe
source ($\stackrel{<}{\sim}$\,11\,\%) than for the pulser calibration
($\stackrel{<}{\sim}$\,15\,\%). This is due to the lower current
demand of the amplifying electronics in the case of the Fe measurement
because only a few pads receive charge signals.

A thorough investigation of cut criteria showed that constraints on
the cluster size in drift direction $r$ and in $\phi$ direction (see
coordinate system in figure~\ref{padrow}) significantly improve the
spectra and reduce noise and background. An effective background
rejection is of great importance for the following calibration with
\nuc{83m}{Kr}. Moreover, clusters located at the marginal pads of each
padrow were rejected.

\subsection{The Krypton Calibration}

In the first step, Krypton spectra were recorded at high anode
voltages (1\,600\,V and 1\,700\,V), focusing on the 9.4\,keV peak. The
calibration factors from the Fe reference measurement were applied to
the data and the spectra from the individual pads were combined to one
spectrum. By this means, the accuracy of the extrapolation of the gas
gain measurements with Fe from high anode voltages to lower voltages
was checked. This extrapolation was correct within an error
$<\,3\,\%$, which includes statistical and systematical errors. The
optimum anode voltage for collecting the complete Krypton spectrum was
determined to be 1\,500\,V.

\begin{figure}
\mbox{\epsfig{file=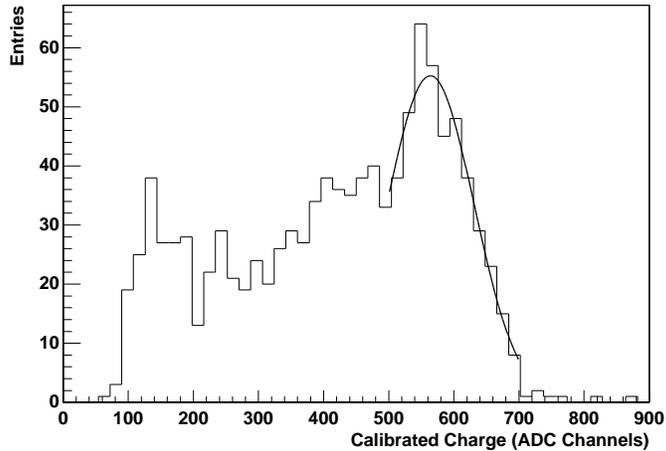,width=0.7\linewidth}}
\hfill
\parbox[b]{.29\textwidth}{\caption{\label{krpad} Krypton spectrum of a
single pad at an anode voltage of 1\,500\,V. At higher anode voltages,
only the low energy part of the spectrum can be detected.}}
\end{figure}

In the second step, a high statistics run at 1\,500\,V was carried out
in order to allow a pad-wise calibration with the 41.6\,keV peak. In
some of the high energy decays, several electrons are produced
consecutively in a short time span. This results in charge clusters
with multiple peaks. As the summed electron energy is relevant, a
special cluster finder\footnote{The cluster finder calculates the
charge and other parameters like position and width of each found
cluster from the raw data.} was used that does not deconvolute
overlapping clusters, but sums up the total charge.

The cuts determined by the Fe measurements were applied and the
calibration constants were then obtained by fitting a simple Gaussian
to the upper part of each spectrum (see figure~\ref{krpad}). This
calibration was repeated iteratively several times since the cluster
charge is summed up from several pads with different calibration
factors although the resulting charge is assigned to the spectrum of
the central pad.

\begin{figure}
\begin{center}
\mbox{\epsfig{file=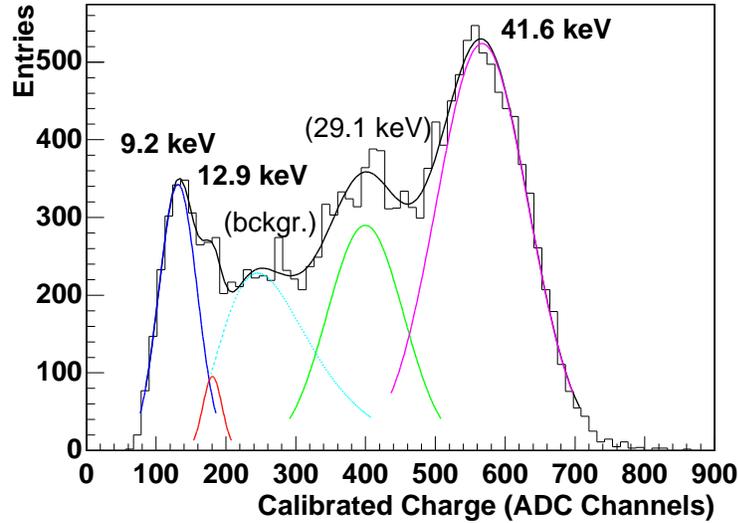,width=0.75\linewidth}}
\parbox{30em}{
\caption{\label{krresult} The measured Krypton spectrum of all pads
calibrated with the calibration constants obtained from the Krypton
spectra of the single pads. A combined fit using 4 Gaussians for the
peaks (solid grey lines) and one Landau distribution for the
background (dotted line) was performed. The maximum of the high energy
peak was assigned an energy of 41.6\,keV and thus the energies of the
other peaks were calculated (see values at the various peaks).}}
\end{center}
\end{figure}

Finally, a total Krypton spectrum was obtained by combining the
calibrated spectra of the single pads (except the marginal pads). This
measured Krypton spectrum is in good agreement with the simulation
(compare figures~\ref{krsim2} and \ref{krresult}). The position of the
peaks were determined by a combined fit to a satisfactory precision.
Both the linearity and the accuracy of the fitting procedure were
checked in the following way: The maximum of the high energy peak was
assigned to an energy of 41.6\,keV. The energies of the other peaks
were then calculated from their fitted positions, assuming
linearity. The deviations of the energies thus obtained from the
expected values were $\approx 2$\,\%. The relative width $\sigma/E$ of the
41.6\,keV peak (i.\,e. the resolution) was determined to be 10\,\%.

The amplification in the gas is sensitive to the density which changes
due to temperature and atmospheric pressure variations. Therefore the
measurements were performed in an air-conditioned room. As the atmospheric
pressure was not monitored during the measurements, a more precise
calibration can be attained when appropriate correction factors are applied
to the data.

In order to obtain a spectrum where the 41.6\,keV peak can be fitted
precisely, at least 1\,000 decay events that fulfill the cut criteria
are required. With moderate cut criteria, 80\,\% of the events pass
the cuts and thus 1\,250 events have to be recorded. Consequently, if
a pad-wise calibration of the FTPC's 19\,200 pads is required, a total
of $24 \cdot 10^6$ events is needed. For a chip-wise calibration (each
chip reads out 16 pads) this number decreases to $1.5 \cdot 10^6$
events. Assuming that the data can be written with a rate of 15\,Hz
and that there is one Krypton decay detected per trigger, it takes
about 28\,hours to record $1.5 \cdot 10^6$ events.

\section{Conclusions}

It was shown that a calibration with Krypton-83m is a very useful and
convenient calibration method for the STAR FTPC. Its practicability is
not affected by the FTPC's pad geometry where the cathode is not
completely covered with pads. With this method, both a compensation of
deviations in amplification within 2\,\% and a precise energy
calibration can be obtained. The energy resolution was determined to
be $\sigma/E=10\,\%$.

\end{document}